\documentclass[twocolumn,10pt,prl,aps,showpacs,superscriptaddress,preprintnumbers,amsmath,amssymb]{revtex4-1}

\usepackage{ragged2e}

\usepackage{graphicx} 
\usepackage{dcolumn} 
\usepackage{bm} 

\usepackage{ulem}
\usepackage[]{color}

\begin{document}

\title{4$f$ crystal field ground state of the strongly correlated topological insulator SmB6$_6$}

\author{M.~Sundermann}
 \affiliation{Institute of Physics II, University of Cologne, Z{\"u}lpicher Stra{\ss}e 77, 50937 Cologne, Germany}
  \affiliation{Max Planck Institute for Chemical Physics of Solids, N{\"o}thnitzer Stra{\ss}e 40, 01187 Dresden, Germany}
\author{H.~Yava\c{s}}
  \affiliation{PETRA III, Deutsches Elektronen-Synchrotron (DESY), Notkestra{\ss}e 85, 22607 Hamburg, Germany}
\author{K.~Chen}
\altaffiliation{present address: Synchrotron SOLEIL, L'Orme des Merisiers, Saint-Aubin, BP~48, 91192 Gif-sur-Yvette C\'edex, France}
\affiliation{Institute of Physics II, University of Cologne, Z{\"u}lpicher Stra{\ss}e 77, 50937 Cologne, Germany}
\author{D.J.~Kim}
  \affiliation{Department of Physics and Astronomy, University of California, Irvine, CA 92697, USA}	
\author{Z.~Fisk}
  \affiliation{Department of Physics and Astronomy, University of California, Irvine, CA 92697, USA}	
\author{D.~Kasinathan}
	\affiliation{Max Planck Institute for Chemical Physics of Solids, N{\"o}thnitzer Stra{\ss}e 40, 01187 Dresden, Germany}
\author{M.~W.~Haverkort}
  \affiliation{Institute for Theoretical Physics, Heidelberg University, Philosophenweg 19, 69120 Heidelberg, Germany}
\author{P.~Thalmeier}
	\affiliation{Max Planck Institute for Chemical Physics of Solids, N{\"o}thnitzer Stra{\ss}e 40, 01187 Dresden, Germany}
\author{A.~Severing}
  \affiliation{Institute of Physics II, University of Cologne, Z{\"u}lpicher Stra{\ss}e 77, 50937 Cologne, Germany}
  \affiliation{Max Planck Institute for Chemical Physics of Solids, N{\"o}thnitzer Stra{\ss}e 40, 01187 Dresden, Germany}
\author{L.~H.~Tjeng}
	\affiliation{Max Planck Institute for Chemical Physics of Solids, N{\"o}thnitzer Stra{\ss}e 40, 01187 Dresden, Germany}

\date{\today}

\begin{abstract}
We investigated the crystal-electric field ground state of the 4$f$ manifold in the strongly correlated topological 
insulator SmB$_6$ using core level non-resonant inelastic x-ray scattering (NIXS). The directional dependence 
of the scattering function that arises from higher multipole transitions establishes unambiguously that the 
$\Gamma_8$ quartet state of the Sm $f^5$ $J$=$5/2$ configuration governs the ground-state symmetry and 
hence the topological properties of SmB$_6$. Our findings contradict the results of density functional calculations reported so far.
\end{abstract}


\maketitle
It was recently proposed that the intermediate valent Kondo insulator SmB$_6$\,\cite{Menth1969,Cohen1970,Allen1979,Gorshunov1999,Riseborough2000} could be a topological insulator\,\cite{Dzero2010,Takimoto2011,Dzero2012,Lu2013,Dzero2013,Alexandrov2013}. Indeed, topologically protected metallic surface states would be an attractive explanation for the low temperature conductance that has been puzzling scientist for decades. The proposal is appealing since in particular rare earth Kondo insulators have the necessary ingredients for strong spin-orbit coupling and electrons of opposite parity, namely the $4f$ and $5d$. The concept of strongly correlated topological insulators is exciting not only because the surface may have massless charge carriers with locked helical spin polarization but also because the surface of such a strongly correlated system may host novel phenomena not present in semiconductor-based topological insulators\,\cite{Vishwanath2013,Bonderson2013,Wang2013,Fidkowski2013}. With the bulk being truly insulating, SmB$_6$ has experienced a tremendous renewed interest and many experimental techniques like angle-resolved photoelectron spectroscopy (ARPES)\,\cite{Xu2013,Zhu2013,Neupane2013,Jiang2013,Denlinger2013, Denlinger2014,Xu2014natcomm, Xu2014, Hlawenka2015}, scanning tunneling spectroscopy\,\cite{Yee2013,Roessler2014,Ruan2014,Roessler2016,Jiao2016}, resistivity and surface conductance measurements\,\cite{Hatnean2013, Zhang2013,Kim2013,Wolgast2013,Kim2014,Wolgast2015,Thomas2016,Nakajima2016} have been applied to unveil its topological properties. Please see also Ref.\,\cite{Dzero2016,AllenReview} and references therein. 

\begin{figure}
    \includegraphics[width=0.99\columnwidth]{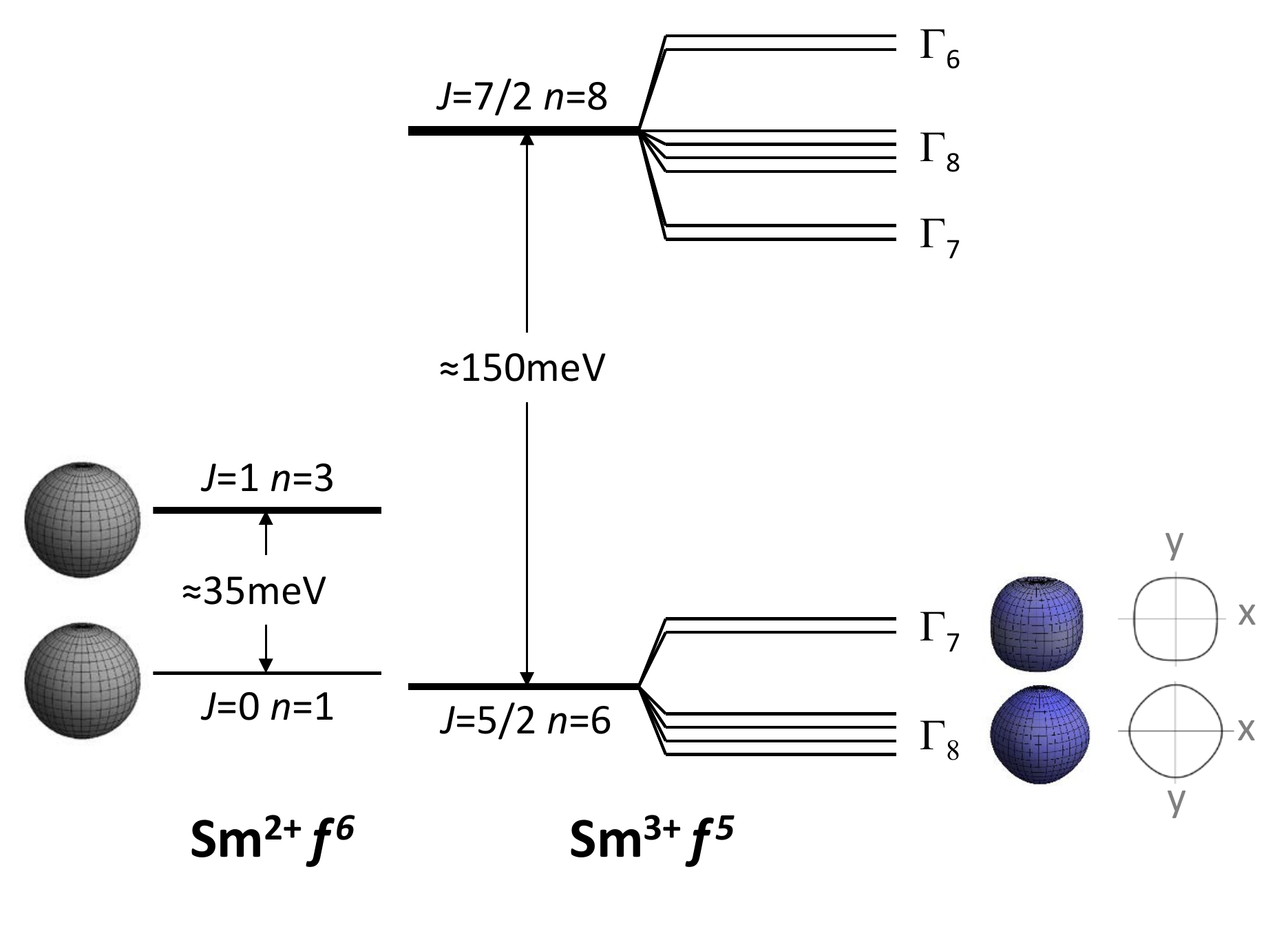}
    \caption{(color online) Sm$^{2+}$ and Sm$^{3+}$ total energy level diagram. The Sm$^{2+}$ 
     configuration is split into a $J$=$0$ and $J$=$1$, and the Sm$^{3+}$ into a $J$=$5/2$ and 
     $J$=$7/2$ multiplet. The label $n$ indicates the degeneracy. The Sm$^{3+}$ multiplets are further split ($\Gamma_i$) by the cubic crystal-electric field. The insets show the corresponding charge densities for six and five electrons and their 2D projections, respectively.}
    \label{fig0}
\end{figure}

In SmB$_6$, the strong hybridization of the low lying 4$f$ states with conduction band $d$ states gives rise to a hybridization gap of the order of 20\,meV\,\cite{Xu2013,Zhu2013,Neupane2013,Jiang2013,Denlinger2013, Denlinger2014}. The Fermi level lies in this hybridization gap so that the material is an insulator when the hybridization becomes effective at low temperatures. The strong hybridization also leads to a partial occupation of the 4$f$ shell or a mixture of the Sm $f^6$ (2+) and $f^5$ (3+) configurations. For the valence at low temperatures, values of 2.5 to 2.7 are given according to different sources in the literature\,\cite{Allen1980,Tarascon1980,Mizumaki2009,Hayashi2013,Lutz2016,Butch2016,Utsumi2017}. Hence the electronic structure is described by the Hund's rule ground states of the Sm $f^6$ (2+) and $f^5$ (3+) configurations with total orbital momenta of $J$\,=\,0 and 5/2, respectively. The $J$\,=\,5/2 multiplet is further split into a $\Gamma_7$ doublet and $\Gamma_8$ quartet due to the cubic crystal-electric field (CEF). Figure\,\ref{fig0} shows the ground state and first excited state of the two Sm configurations plus their electron charge density distributions. The charge densities of the $J$\,=\,0 and 1 states are spherical since neither the $J$\,=\,0 or 1 are split in a cubic potential\,\cite{Lea1962}. This is contrasted by the charge densities of the CEF split $J$\,=\,5/2 multiplet (and $J$\,=\,7/2, not shown) that are anisotropic.

Information of the surface topology can be unambiguously inferred from the symmetries and parities of the bulk states involved. Knowledge about the CEF ground state symmetry of SmB$_6$ therefore plays an essential role. For example, theoretical predictions for the spin texture of the sought-after topological surface states depend very much whether the ground state of the $f^5$ $J$=$5/2$ configuration is the $\Gamma_8$ quartet or the $\Gamma_7$ doublet CEF state\,\cite{Baruselli2015,Legner2015,Baruselli2016}.

Surprisingly, after forty years of research, the CEF scheme of SmB$_6$ has still to be determined. The classical tool inelastic neutron scattering has not been able to identify the CEF states, possibly due to the superposition of both Sm $f^5$ and $f^6$ configurations in this mixed valent compound and the strong neutron absorption despite double isotope samples\,\cite{Alekseev1993,Alekseev1995,Fuhrman2015}. From inelastic neutron scattering, a sharp excitation at 14\,meV close to the hybridization gap was reported. It was assigned to a spin exciton and not to a CEF excitation since its intensity does not follow the 4$f$ magnetic form factor. Further magnetic intensities at about 35\,meV, 115\,meV, and 85\,meV have been assigned to the inter-multiplet transitions of the Sm$^{2+}$ configuration and of the CEF split Sm$^{3+}$ configuration (see Fig.\ref{fig0}), and to some magnetoelastic coupling, respectively. In-gap transitions at about 15\,meV in Raman spectra could be interpreted as CEF excitations but Raman does not yield the information about which state forms the ground state\,\cite{Nyhus1995,Nyhus1997}. A semi-empirical extrapolation method can predict CEF parameters across the rare earth series for highly diluted systems\,\cite{Frick1986}. Applying such an extrapolation to the measured CEF schemes of REB$_6$ with RE\,=\,Ce, Pr and Nd\,\cite{Zirn1984,Loewenhaupt1986} yields for SmB$_6$ a CEF splitting of the order of 15\,meV with the $\Gamma_8$ quartet as the ground state. However, the Kondo insulator SmB$_6$ is not a highly diluted system and it is definitely not an ionic system but highly intermediate valent instead, questioning the validity of such an extrapolation.

We, therefore, performed bulk-sensitive, core-level non-resonant inelastic hard-x ray scattering (NIXS) measurements that target specifically the ground state symmetry of SmB$_6$. NIXS is a powerful tool to determine the ground state wave function of $4f$ and 5$f$ systems\,\cite{Willers2012,Rueff2015,Sundermann2015,Sundermann2016}. This bulk sensitive and element specific spectroscopic method is carried out with large momentum transfers $|\vec{q}|$ so that the transition operator e$^{i\vec{q}\cdot\vec{r}}$ in the scattering function S($\vec{q}$,$\omega$) contains contributions of higher multipole terms, giving information that is not accessible in a dipole experiment\,\cite{Soininen2005,Larson2007,Haverkort2007,Gordon2008,Gordon2009,Bradley2010,Caciuffo2010,Rueff2010,Hiraoka2011,SenGupta2011,Bradley2011,Laan2012}. Here, the dependence of S($\vec{q}$,$\omega$) on the direction of vector $\vec{q}$ with respect to the crystallographic lattice provides the symmetry information of the ground state wave function, even for cubic 
compounds thanks to the multipole terms\,\cite{Gordon2009,Sundermann2017}. 

The NIXS measurements on the Sm and Eu \,$N_{4,5}$ core level ($4d^{10}4f^5$\,$\rightarrow$\,$4d^{9}4f^6$ and $4d^{10}4f^6$\,$\rightarrow$\,$4d^{9}4f^7$, respectively) were performed at the beamline P01 of PETRA-III with a fixed final energy of 9690\,eV, an energy resolution of $\approx$\,0.7\,eV full width at half maximum (FWHM), and an averaged momentum transfer of $|\vec{q}|$\,=\,(9.6\,$\pm$\,0.1)\,\AA$^{-1}$. Further experimental details can be found in Supplementary Material\,\cite{Sup}. 

The SmB$_6$ single crystals were grown by the aluminum flux method \cite{Kim2014}, the polycrystalline commercial reference samples Sm$_2$O$_3$ ($4f^5$) and Eu$_2$O$_3$ ($4f^6$) were pressed pellets with a purity of 99.9\,\% and 99.99\,\%, respectively. All samples were mounted in a vacuum cryostat with Kapton windows and measured at 16\,K. Two SmB$_6$ single crystals with (100) and (110) surfaces were oriented such that for $\vec{q}$\,$\|$\,[100] and $\vec{q}$\,$\|$\,[110] a specular scattering geometry was realized. For the $\vec{q}$\,$\|$\,[111] direction one of the crystals was turned accordingly with respect to the scattering triangle. 

\begin{figure}
    \includegraphics[width=0.99\columnwidth]{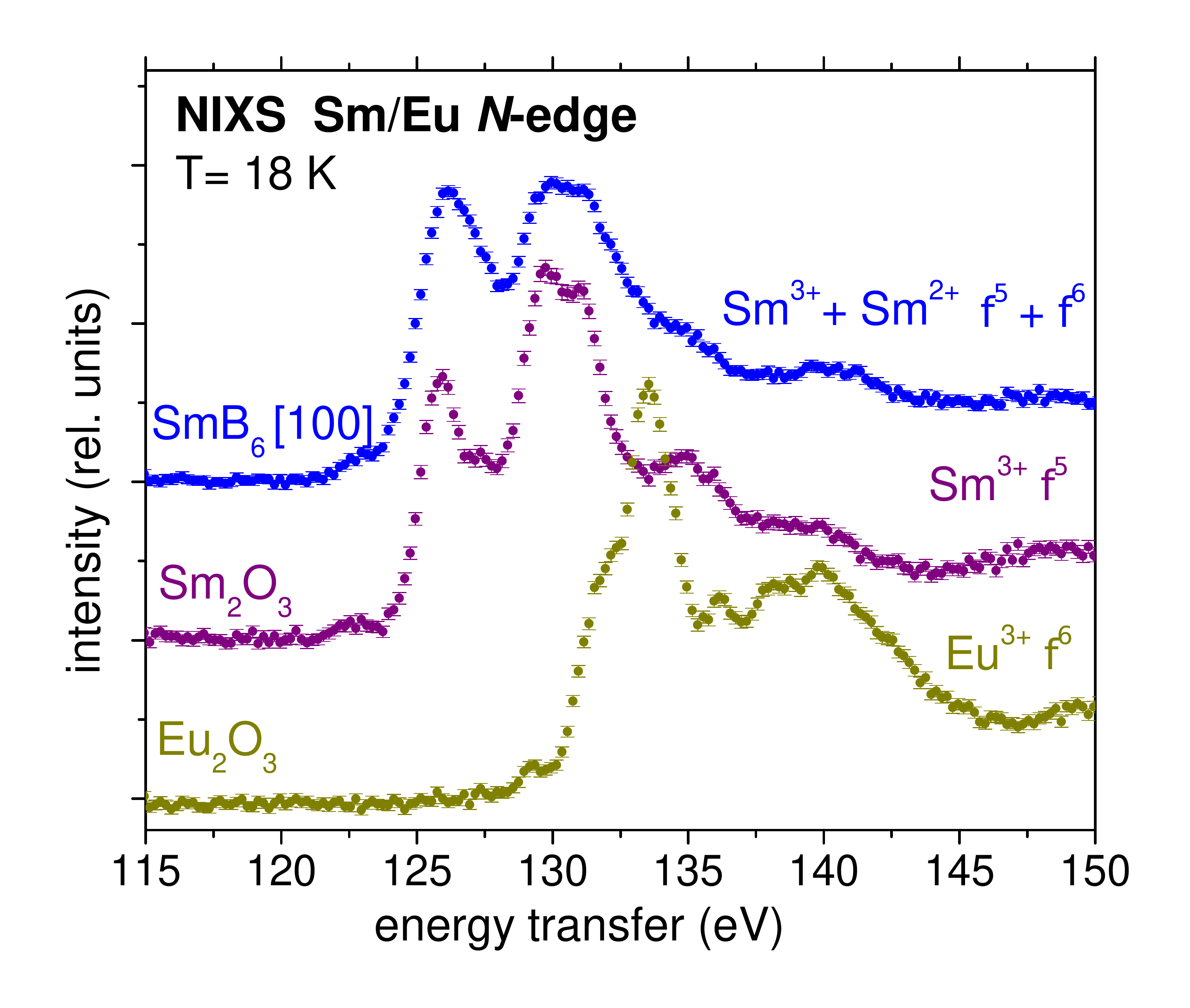}
    \caption{(color online) Energy scans at the $N_{4,5}$ edges of SmB$_6$, Sm$_2$O$_3$ and Eu$_2$O$_3$
    after subtracting a linear background.}
    \label{fig2}
\end{figure}

Fig.\,\ref{fig2}\ shows the NIXS spectra across the $N_{4,5}$ edges of SmB$_6$ (blue dots) and of the two reference compounds Sm$_2$O$_3$ and Eu$_2$O$_3$  (purple and dark yellow) after subtraction of a linear background and scaling to the Compton background. Spectra over a larger energy interval showing also the elastic lines and the Compton background are given in Fig.\,S1 of the Supplementary Material\,\cite{Sup}. The Eu edge appears at a higher energy transfer than in the case of Sm because Eu has a higher atomic number. 

\begin{figure}
    \includegraphics[width=0.99\columnwidth]{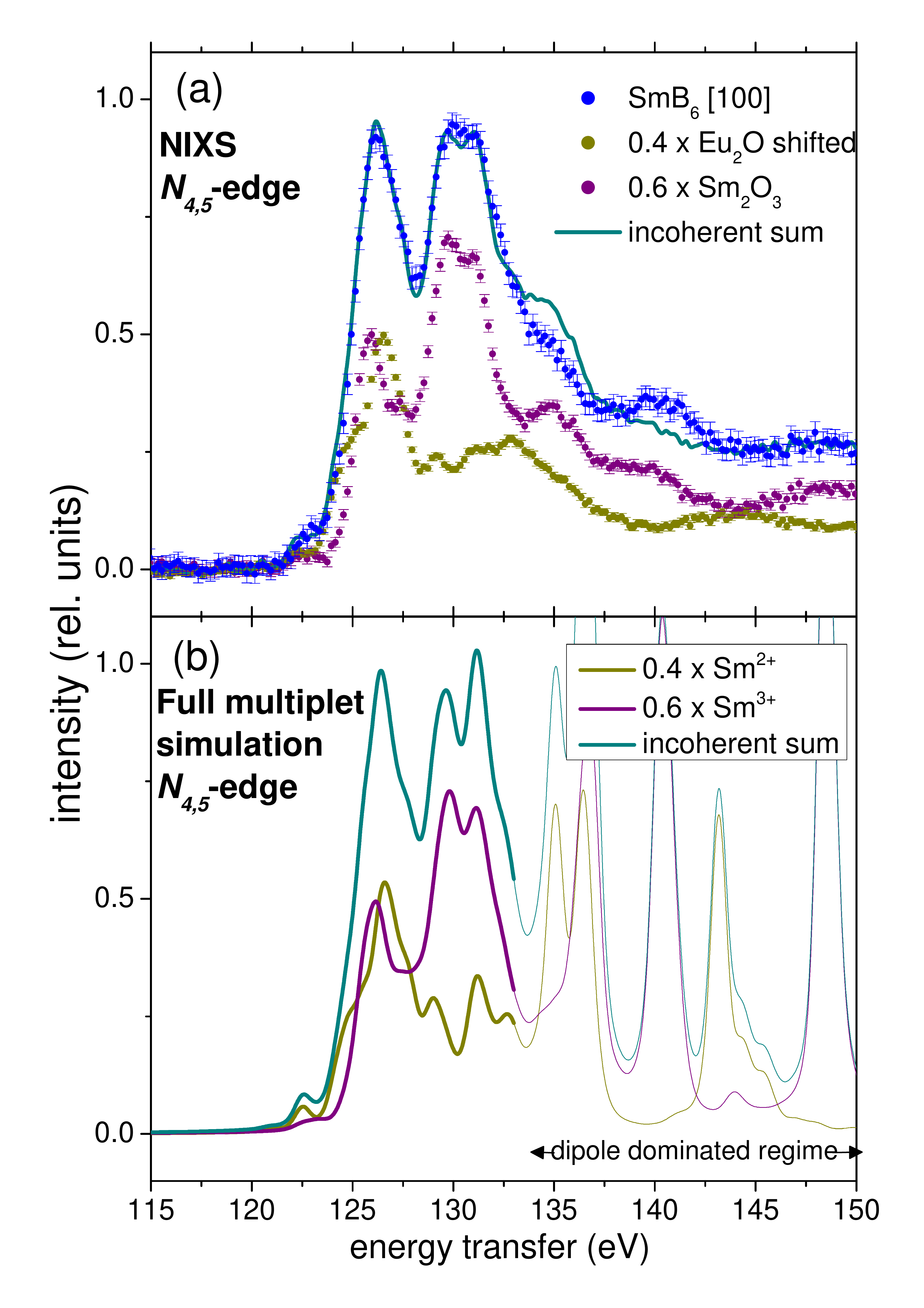}
    \caption{(color online) (a) Experimental SmB$_6$ data for $\vec{q}$$\|$[100] (blue dots) together with
    the weighted sum (dark cyan line) of the experimental Sm$_2$O$_3$ ($f^5$) (purple dots) and energy shifted 
    experimental Eu$_2$O$_3$ ($f^6$) (dark yellow dots). (b) Full multiplet simulation of Sm$^{3+}$ (purple line) and Sm$^{2+}$ 
    spectra (dark yellow line) and their weighted sum (dark cyan line).}
    \label{fig3}
\end{figure}

We first investigate whether the SmB$_6$ spectrum can be interpreted using those of Sm$_2$O$_3$ and Eu$_2$O$_3$. For this purpose we construct a spectrum made up of the weighted sum of Sm$_2$O$_3$ and Eu$_2$O$_2$. The best reproduction of the data is obtained weighing the Sm reference with a factor 0.6 and the Eu data with a factor of 0.4. In addition, the Eu$_2$O$_3$ spectrum is shifted by 6.8\,eV to lower energies in order to account for the higher atomic number. The resulting spectrum reproduces the SmB$_6$ spectrum very satisfactorily (see dark cyan line in Fig.\,\ref{fig3}\,(a)). The weights used for the sum correspond to a Sm valence of 2.6, in good agreement with other studies using a variety of different experimental methods\,\cite{Allen1980,Tarascon1980,Mizumaki2009,Hayashi2013,Lutz2016,Butch2016,Utsumi2017}. This provides us with confidence to carry out further analysis using full multiplet calculations based on the 4$f^5$ and 4$f^6$ configurations of Sm.

Fig.\,\ref{fig3}\,(b) shows the full multiplet simulation of the Sm$^{3+}$ $N_{4,5}$ edges (purple line) resulting from a fit to the Sm$_2$O$_3$ data (see Fig.\,S2 in Supplementary Material\,\cite{Sup}). The $N_{4,5}$ edge of Sm$^{2+}$ (dark yellow line) was calculated using the same adjustable parameters as for Sm$^{3+}$ (see below). The weighted sum (60\% and 40\%) of the simulated curves (dark cyan) describes the SmB$_6$ spectrum very well in the energy region between 120 and 135 eV. This is the region where the high multipole scattering dominates (see Fig.\,S3 of the Supplementary Material\,\cite{Sup} and Ref.\cite{Gordon2008} for further explanation). In the region above $\approx$\,135\,eV, where the spectrum is given mostly by the dipole transitions (see Fig.\,S3 and Ref.\cite{Gordon2008}) the simulation produces spectral features that are too sharp with respect to the experiment because the interference with the continuum states is not included in the calculations. The high multipole excitations are more realistically reproduced since they are lower in energy and therefore further away from the continuum states and consequently more excitonic\,\cite{Gupta2011}.

The $4d$\,$\rightarrow$\,$4f$ transitions were simulated with the full multiplet code \textsl{Quanty} which includes Coulomb and spin-orbit interactions\,\cite{Haverkort2016}. A Gaussian and a Lorentzian broadening of 0.7\,eV and 0.4\,eV FWHM, respectively, account for the instrumental resolution and life-time effects. The atomic 4$f$-4$f$ and 4$d$-4$f$ Coulomb interactions were calculated using the Hartree-Fock scheme and a reduction of about 20\,\%\,\footnote{The 4$d$-4$f$ Slater integral G1 has been reduced another 15\% to about 68\%.} has been applied to obtain the best agreement between the calculated and measured peak positions\,\cite{Tanaka1994}. Further details about the simulation can be found in Supplementary Materials\,\cite{Sup}.

\begin{figure}
    \includegraphics[width=0.99\columnwidth]{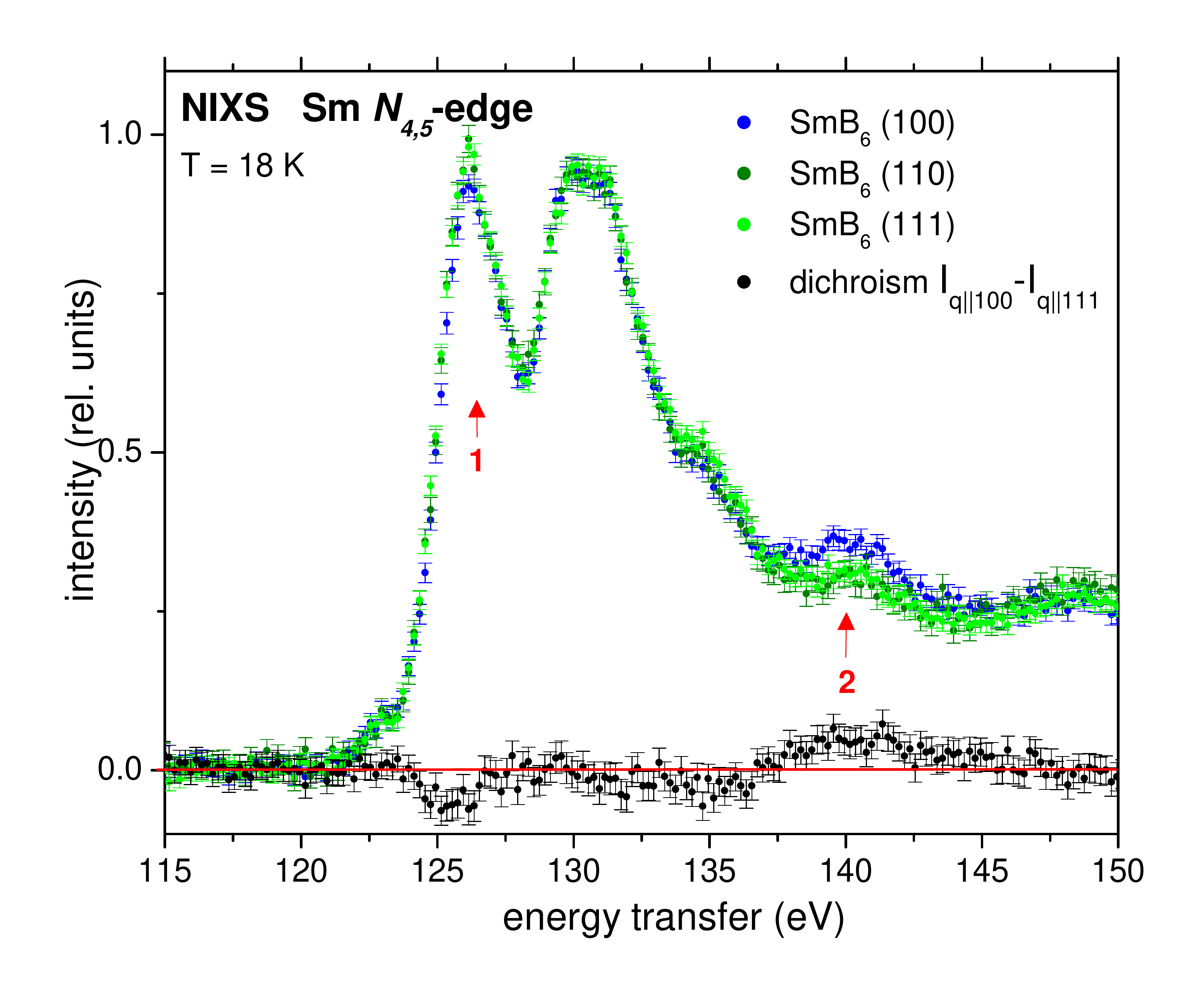}
    \caption{(color online) SmB$_6$ NIXS data at 16\,K for $\vec{q}$$\|$[100] (blue dots), 
    $\vec{q}$$\|$[110] (dark green dots), and $\vec{q}$$\|$[111] (light green dots). 
     The difference spectrum between the $\vec{q}$$\|$[100] and $\vec{q}$$\|$[111] directions
     is also displayed (black dots).}
    \label{fig4}
\end{figure}

Figure\,\ref{fig4} shows the direction dependence of the Sm $N_{4,5}$ of SmB$_6$. Although the effect is small, there are clear differences between the spectra in the energy regions marked with red arrows.  At about 126\,eV energy transfer the scattering of the $\vec{q}$$\|$[110] (light green dots) and $\vec{q}$$\|$[111] (dark green dots) directions are both stronger than for the $\vec{q}$$\|$[100] (blue dots), and at about 140\,eV it is opposite. To show these directional differences in a more transparent manner, we also present in Fig.\,\ref{fig4} the difference spectrum between the $\vec{q}$$\|$[100] and $\vec{q}$$\|$[111] (black dots): this so-called dichroic spectrum has unambiguously a negative peak at 126 eV whereas it displays positive intensity in a broader region around 140\,eV.

\begin{figure}
    \includegraphics[width=0.99\columnwidth]{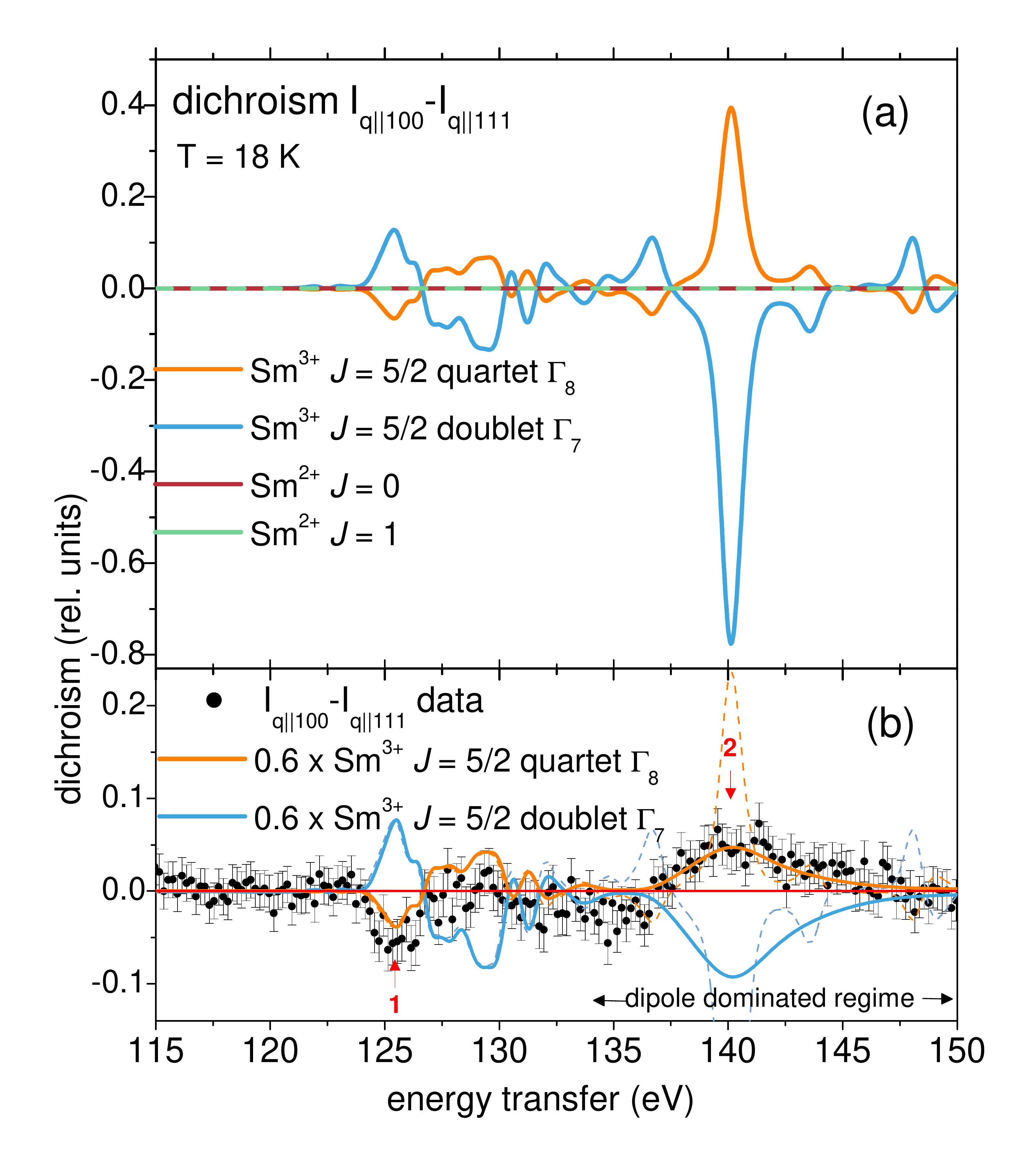}
    \caption{(color online) (a) simulation of  the $\vec{q}$$\|$[100] vs. $\vec{q}$$\|$[111] dichroic spectrum 
     for the $J$=$0$ (brown) and $J$=$1$ (green) multiplet states of the Sm$^{2+}$ configuration as well as
     for the $\Gamma_8$ quartet (orange) and $\Gamma_7$ doublet (light blue) of the $J$=$5/2$ Sm$^{3+}$ 
     configuration;
    (b) experimental dichroic spectrum (black dots) and simulated dichroic spectra for the $\Gamma_8$ quartet 
     (orange) and $\Gamma_7$ doublet (light blue) scaled with the factor of 0.6 to account for the Sm$^{3+}$ 
     component of the ground state; dashed lines with energy independent broadening, solid lines with extra broadening in the dipole region (see text).}
    \label{fig5}
\end{figure}

To interpret the observed direction dependence, it is important to know how each CEF state or multiplet 
component contributes to the dichroic signal. Therefore, S($\vec{q}$,$\omega$) has been calculated 
taking into account a cubic CEF for the Sm$^{3+}$ $f^5$ ground state multiplet with $J$=$5/2$ assuming 
a $\Gamma_8$ quartet or a $\Gamma_7$ doublet ground state, and for the Sm$^{2+}$ $f^6$ multiplets 
with $J$=$0$ or $J$=$1$ (see Fig.\ref{fig0}),\,\footnote{The excited $J$\,=\,7/2 multiplet is not considered 
because it will not contribute since it is too high in energy.}. The calculations were  performed for the two 
directions $\vec{q}$$\|$[100] and $\vec{q}$$\|$[111] and in Fig.\,\ref{fig5}\,(a) the resulting dichroic 
signals are plotted. 
The calculated dichroism for the [110] and [111] direction is shown in the Supplementary Materials Fig.\,S4. 
Here only the multipole scattering contributes to the dichroism, the dipole does not because the Sm 
site symmetry is cubic.

The first important finding is that the Sm$^{2+}$ configuration does not show any dichroism at all (see dark red and green lines at zero dichroism) as we would expect for states with spherical charge densities (see Fig.\,\ref{fig0} and \textsl{Direction Dependence in NIXS} in Supplementary Materials, which includes Refs.\,\cite{Csiszar2005,Haverkort2004}). Hence, the observed direction dependence of the signal is solely due to the initial state of the Sm$^{3+}$ Hund's rule ground state. The second important finding is that the $\Gamma_8$ and $\Gamma_7$ CEF states exhibit different and opposite dichroism (see orange and light blue lines), consistent with their opposite anisotropy in the charge densities (see Fig.\,\ref{fig0}). The opposite dichroism at 125 and 140\,eV reduces the experimental challenge to simple yes/no experiment and makes the determination of the CEF ground state of Sm$^{3+}$ in SmB$_6$ straightforward.

Figure\,\ref{fig5}\,(b) shows the experimental dichroic spectrum (black dots) together with the calculated ones. The two possible CEF states of the $J$=$5/2$ configuration have now been scaled down to 60\% to quantitatively account for the Sm$^{3+}$ component in intermediate valent SmB$_6$. We can clearly observe that in the regions of pronounced dichroism (see red arrows) the sign of the experimental dichroic signal is correctly explained by the $\Gamma_8$ quartet (orange line) but not at all by the $\Gamma_7$ 
doublet state (light blue line). In addition, the $\Gamma_8$ reproduces the experimental dichroism quantitatively in the high multipole region (see red arrow 1). The dichroism also fits quantitatively in the dipole region (see red arrow 2) when an extra broadening is applied (FWHM $\geq$ 4\,eV beyond $\approx$135\,eV energy transfer) to mimic the interference with continuum states. Note that sum rules still apply, i.e. the interference with the continuum states does not change the polarization, it only affects the broadening. The dashed lines correspond to the dipole calculation without the extra broadening. These results unambiguously establish that the CEF ground state of the Sm $f^5$ component in SmB$_6$ is the $\Gamma_8$ quartet. 

We would like to point out that the down scaling to 60\% of the Sm $f^5$ component gives a good quantitative agreement in the magnitude of the dichroic signal. This provides confidence that the NIXS method is indeed reliable since this 60\% number is fully consistent with the existing valence determination in the literature \cite{Allen1980,Tarascon1980,Mizumaki2009,Hayashi2013,Lutz2016,Butch2016,Utsumi2017} as well as with the above analysis of the total N$_{4,5}$ NIXS spectra. We also would like to note that possible errors in the alignment of the Sm$^{2+}$ NIXS signal with respect to that of the Sm$^{3+}$ do not affect the dichroic signal and hence the analysis of the CEF ground state since the Sm$^{2+}$ is silent in terms of directional dependence.

Our finding of the $\Gamma_8$ quartet forming the ground state supports very much the results of spin resolved APRES\,\cite{Xu2014natcomm}.  Xu \textsl{et al.} find spin polarized surface states, fulfilling time reversal as well as crystal symmetry, that have spins locked to the crystal momenta $k$ such that at opposite momenta the surface states have opposite spins. The anticlockwise spin texture is in agreement with spin expectation values that are calculated by Baruselli and Vojta for a $\Gamma_8$ ground state\,\cite{Baruselli2015,Baruselli2016}. Note, for a $\Gamma_7$ the spin directions should be reversed. 
 
Our finding of a $\Gamma_8$ local ground-state symmetry contradicts the outcome of several density 
functional band structure calculations\,\cite{Yanase1992,Antonov2002,Lu2013,Kang2015}. 
 In band theory, the search for the ground state symmetry in SmB$_6$ translates into the question
in which band the hole in the J\,=\,5/2 manifold resides.
Kang \textsl{et al}. reported for the $X$-point an unoccupied $4f$ state of $\Gamma_7$ origin\,\cite{Kang2015}. 
Also their $k$-integrated $4f$ $J=5/2$ partial density of states (pDOS) shows the hole residing in the 
$\Gamma_7$ band, in line with the fact that the center of gravity of the $\Gamma_7$ pDOS is higher 
in energy than that of the $\Gamma_8$, and despite the fact that the $\Gamma_7$ band is lower than 
the $\Gamma_8$ at the $\Gamma$ point. Our experiments showed instead that the $X_7^-$ band at the X-point that is above the Fermi level originates from the $\Gamma_8$ and 
not from the $\Gamma_7$.

To summarize, we have utilized the high multipole contributions in the core-level non-resonant inelastic x-ray scattering process to determine the symmetry of the Sm crystal field ground state 4$f$ wave function in SmB$_6$. We have found a clear directional dependence of the spectra that allows for the unambiguous identification of the $\Gamma_8$ quartet state of the Sm $f^5$  $J$=$5/2$ configuration as the state which governs the topological properties of SmB$_6$. Follow-up calculations should be performed within a reduced basis of only $\Gamma_8$  states for the construction of a low-energy many-body Hamiltonian.

\section{Acknowledgment}K.C., M.S.,A.S., and D.K. benefited from the financial support of the Deutsche Forschungsgemeinschaft under projects SE\,1441 and SPP\,1666. Parts of this research were carried out at PETRA III at DESY, a member of the Helmholtz Association (HGF). We thank C.~Becker and T.~Mende from MPI-CPfS, and F.-U.~Dill, S.~Mayer, H.C. Wille, and M.~Harder from PETRA III at DESY  for their skillful technical support. 

%

\section{Supplementary Material}

\subsection{Experimental and Simulation}

At beamline P01 of PETRA-III the incident energy is selected with a Si(311) double monochromator. The P01 NIXS end station has a vertical geometry with twelve Si(660) 1\,m radius spherically bent crystal analyzers that are arranged in 3\,x\,4  array as shown in Fig.\,2 of Ref.\,\onlinecite{Sundermann2017}. The fixed final energy was 9690\,eV. The analyzers were positioned at scattering angles of 2\,$\theta$\,$\approx$\,150$^\circ$, 155$^\circ$, and 160$^\circ$ which provide an averaged momentum transfer of $|\vec{q}|$\,=\,(9.6\,$\pm$\,0.1)\,\AA$^{-1}$. The scattered beam was detected by a position sensitive custom-made Lambda detector, based on a Medipix3 chip detector. The elastic line was consistently measured and a pixel wise calibration yields an instrumental energy resolution of $\approx$\,0.7\,eV full width at half maximum (FWHM).

The reduction of the Slater integrals in the full multiplet simulation accounts for configuration interaction processes not included in the Hartree-Fock scheme\,\cite{Tanaka1994}. A momentum transfer of $|\vec{q}|$\,=\,9.8\,\AA$^{-1}$ has been used for the simulations rather than the experimental value of 9.6\,$\pm$\,0.1\,\AA$^{-1}$ in order to reproduce best the experimental peak intensity ratio of the two main features around 126 and 130\,eV. This fine tuning optimizes the different multipole contributions to the scattering functions as shown in Fig.\,S3 and represents a minor adjustment of the calculated radial wave functions of the Sm$^{3+}$ atomic wave function (see e.g.\ Ref.\,\cite{Willers2012}).

\subsection{Compton Scans}

\begin{figure}
    \includegraphics[width=0.95\columnwidth]{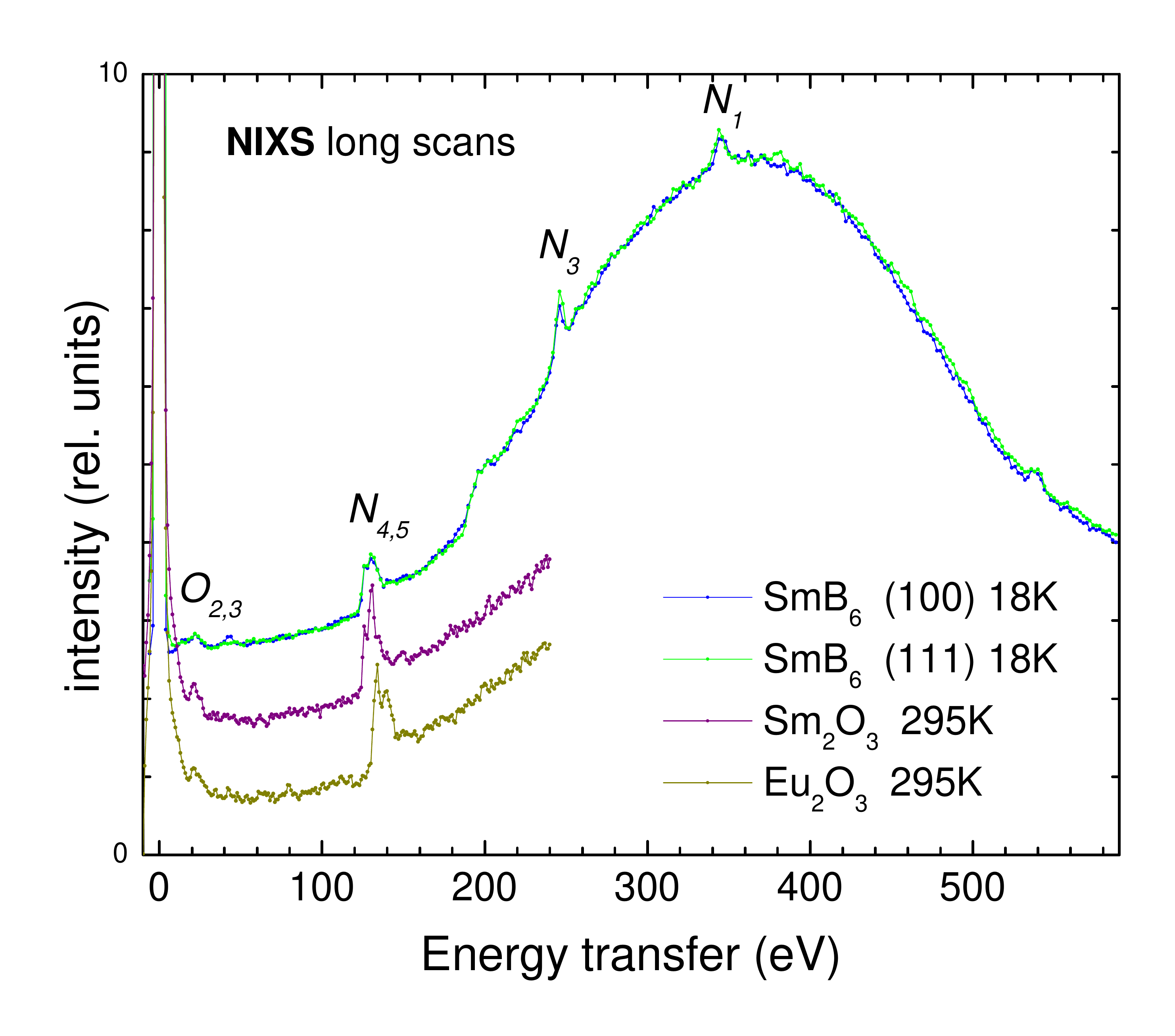}
    \justify
    FIG.\,S1\,\,(color online) Energy scans of SmB$_6$ for $\vec{q}$$\|$[100] and $\vec{q}$$\|$[111], 
    Sm$_2$O$_3$ and Eu$_2$O$_3$ measured at $|\vec{q}|$\,=\,(9.6\,$\pm$\,0.1)\,\AA$^{-1}$ and with  
		a constant final energy of 9690\,eV over a wide energy ranges, for SmB$_6$ with 0.5\,eV energy steps, 
    for Sm$_2$O$_3$ and Eu$_2$O$_3$ with 0.2\,eV; shifted for clarity. 
    \label{}
\end{figure}

Fig.\,S1 shows non-resonant inelastic x-ray scattering (NIXS) spectra of SmB$_6$ (blue and green line) 
measured over a large energy interval and of the two reference compounds Sm$_2$O$_3$ and Eu$_2$O$_3$
(purple and dark yellow) measured up to 250\,eV energy transfer. The strongest signal is the elastic line, 
followed by shallow Sm/Eu core resonances (O-edges) and the Sm/Eu N$_{4,5}$-edges at about 130 and 
135\,eV sitting on top of the rising Compton background. The spectra are offset on the y-axis by 1 unit 
after scaling to the Compton background. 

\subsection{$N$-edge data of S{m}$_2$O$_2$}

\begin{figure}
    \includegraphics[width=\columnwidth]{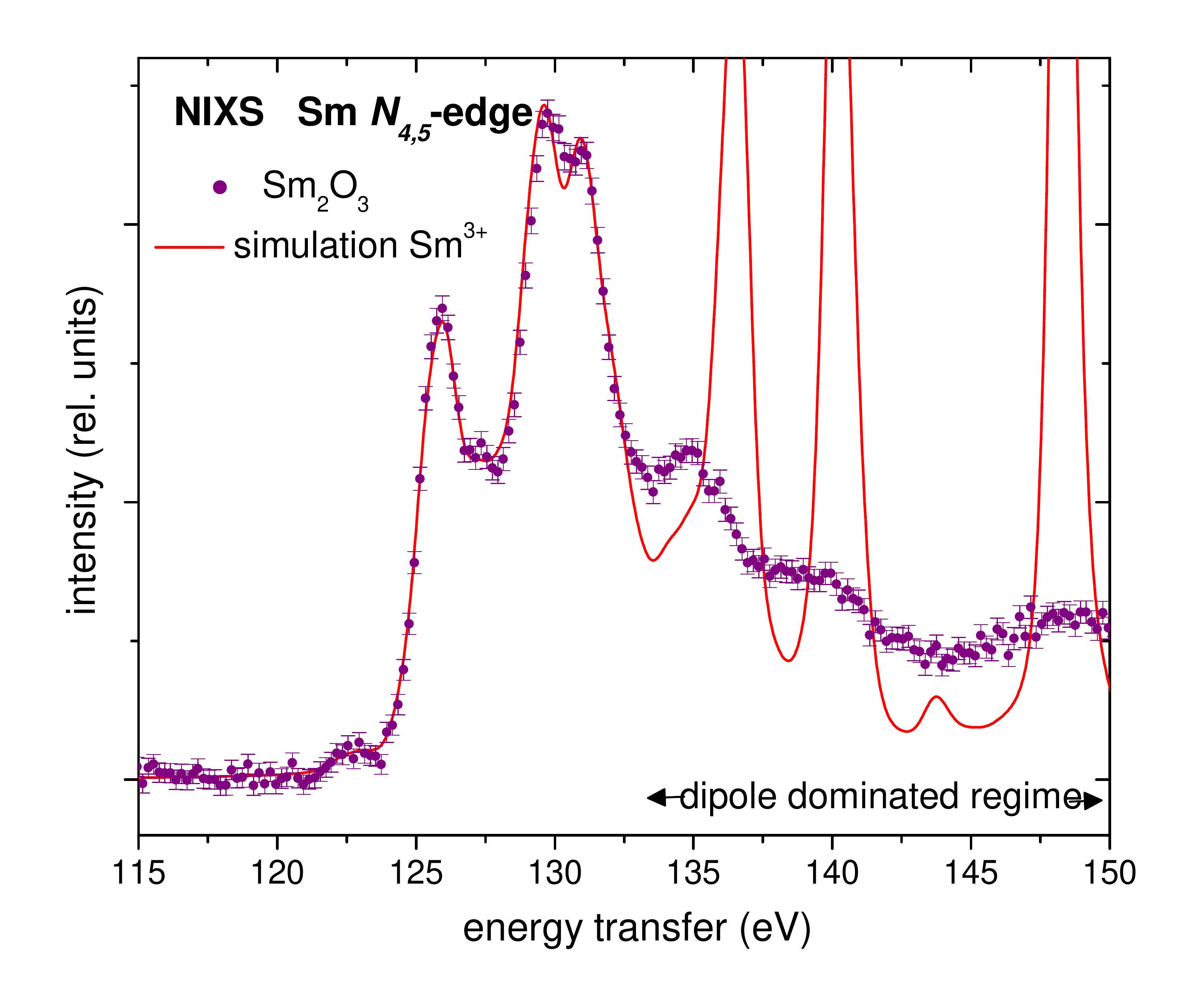}
    \justify
    FIG.\,S2\,\,(color online) $N$-edge data of Sm$_2$O$_2$ (purple dots) measured with 
    $|\vec{q}|$\,=\,(9.6\,$\pm$\,0.1)\,\AA$^{-1}$ for a constant final energy of 9690\,eV 
    and simulation (red line).
    \label{}
\end{figure}

Fig.\,S2 shows the fit (red line) of the Sm$_2$O$_3$ NIXS data (purple dots) at the Sm 
$N_{4,5}$-edge using the full multiplet code $Quanty$\,\cite{Haverkort2016}. In the multipole 
region the agreement between calculation and data is good, in the dipole region the 
simulation yields unrealistically sharp features since interference with continuum states are 
not included in the calculation.

\subsection{Momentum and Energy Dependence of Scattering Function}

\begin{figure*}
    \includegraphics[width=1.9\columnwidth]{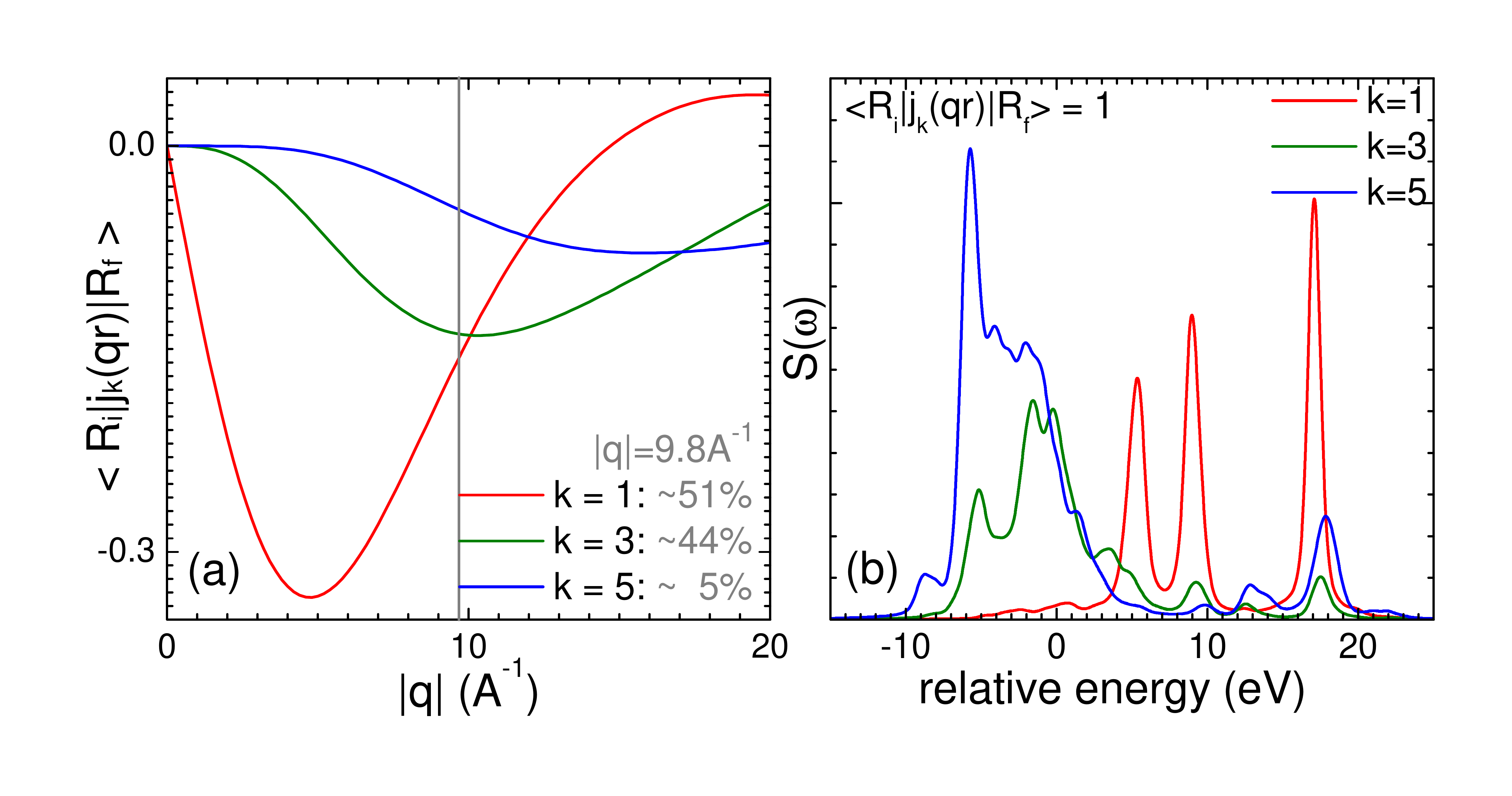}
    \justify
    FIG.\,S3\,\,(color online) Momentum $|\vec{q}|$ dependence (left) and energy dependence 
    (right) of the scattering function S($\vec{q}$,$\omega$) at the Sm $N_{4,5}$-edge for 
    dipole (k=1), octupole (k=3), and dotriacontapole (k=5) scattering orders. The grey vertical 
    line marks the $\left|\vec{q}\right|$-range of the experiment.
    \label{}
\end{figure*}

In scattering experiments at large momentum transfers $|\vec{q}|$ the scattering function 
S($\vec{q}$,$\omega$) contains higher than dipole terms. Fig.\,S3\,(a) shows the radial 
part of  S($\vec{q}$,$\omega$) for dipole (k\,=\,1 red line) and higher orders 
(k\,=3, and 5; green and blue lines). Already at $|\vec{q}|$\,$\cong$\,10\,\AA$^{-1}$ 
the higher order terms amount to about 50\,\% of the total scattering intensity. 
Note, 10\,\AA$^{-1}$ is accessible in a hard x-ray NIXS experiment. The excitations due to 
k\,=\,3\,and\,5 appear lower in energy than the dipole excitations (see fig.\,S3\,(b)). 
Not only the sensitivity of beyond dipole scattering to higher than two-fold symmetries 
is an advantage, in addition these beyond dipole excitations are more excitonic 
(see e.g. in Ref.\,\cite{Gordon2008,Caciuffo2010}). The method is bulk sensitive and 
can be modeled quantitatively (see e.g. Ref.\,\cite{Willers2012,Rueff2015,Sundermann2016,Sundermann2017}).

\subsection{Direction Dependence in NIXS}
Direction dependence in NIXS (and analogously, linear-polarization dependence in XAS) can originate from a different orbital occupation in the initial state (see e.g. Ref. \cite{Csiszar2005}]) and/or due to different energy levels of the orbitals in the final state (see e.g. Ref. \cite{Haverkort2004}). In SmB$_6$ the $f^6$ initial state has a spherical charge density, yielding no contribution to direction dependence. The final state crystal-field induced splitting in Sm is negligible compared to the inverse lifetime of the core hole, resulting in vanishingly small direction dependence.

Fig.\,S4(a) and (b) compare the calculated difference of the scattering intensities for $\vec{q}$$\|$[100] and $\vec{q}$$\|$[111], and $\vec{q}$$\|$[110] and $\vec{q}$$\|$[111]. The dichroic signal is small. It should be kept in mind that the Sm $N_{4,5}$ edge is a scattering process from 4$d$\,$\rightarrow$\,4$f$, i.e. from a state with a complicated shaped charge distribution to another one with a non-spherical shape.  It is therefore hard to predict the scattering intensity as function of direction from intuition. It rather requires a full multiplet calculation taking initial and final states into account.

\begin{figure*}
    \includegraphics[width=1.9\columnwidth]{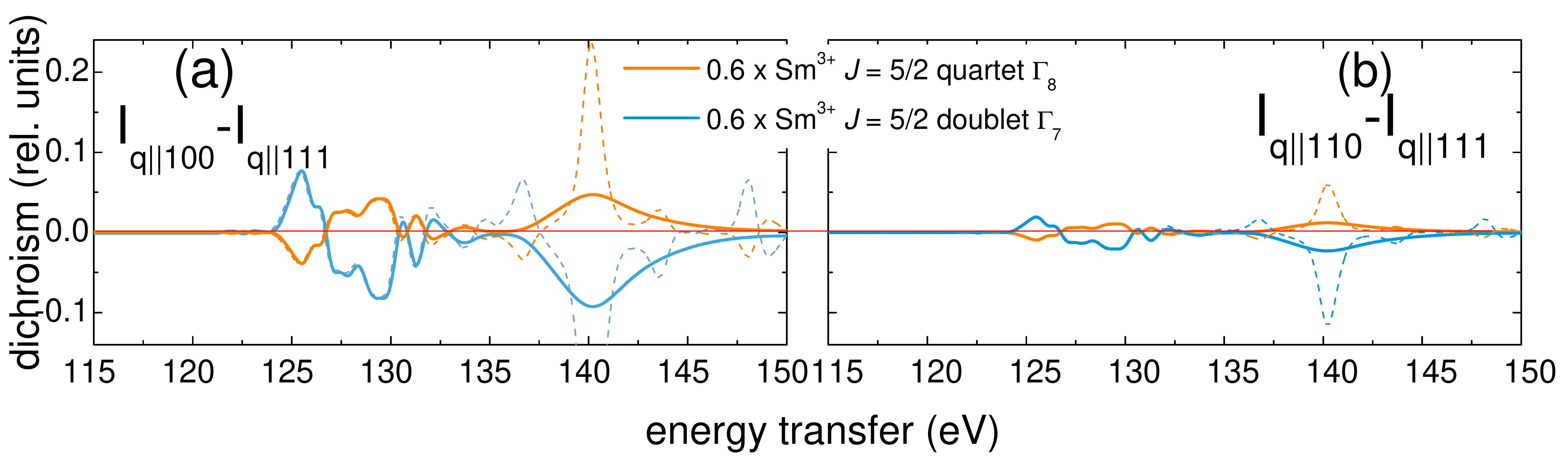}
    \justify
    FIG.\,S4\,\,(color online) Calculated dichroic spectra for the $\Gamma_8$ quartet (orange) and $\Gamma_7$ doublet (light blue) scaled with the factor of 0.6 to account for the Sm$^{3+}$ component of the ground state: (a)\,$\vec{q}$$\|$[100] vs. $\vec{q}$$\|$[111] as in main manuscript (Fig. 5b) and (b)\,$\vec{q}$$\|$[110] vs. $\vec{q}$$\|$[111]; dashed lines with energy independent broadening, solid lines with extra broadening in the dipole region.
    \label{}
\end{figure*}


%

\end{document}